\documentclass[10pt]{article}
\usepackage{sao2}
\usepackage{psfig,epsf}

\issue{2009, 64, 287--296}

\begin{document}
\markboth{M.L.~Khabibullina, O.V.~Verkhodanov}
     {CATALOG OF RADIO GALAXIES WITH $z>0.3$. II: Photometric Data}
\title{Catalog of Radio Galaxies with~ $z>0.3$. II: Photometric Data}
\author{M.L.~Khabibullina\inst{a}
\and O.V.~Verkhodanov\inst{a}
}
\institute{
\saoname
}
\date{December 1, 2008}{April 4, 2009}
\maketitle

\begin{abstract}
We describe the procedure of the construction of a sample of
distant ($z>0.3$) radio galaxies using the NED, SDSS, and CATS
databases. We believe the sample to be free of objects with quasar
properties. This paper is the second part of the description of
the radio galaxies catalog we plan to use for cosmological tests.
We report the photometric parameters for the objects of the list,
and perform its preliminary statistical analysis including the
construction of the Hubble diagrams.
\mbox{\hspace{1cm}}
\end{abstract}

\section{INTRODUCTION}
\begin{figure*}[tbp]
\centerline{\hbox{
\psfig{figure=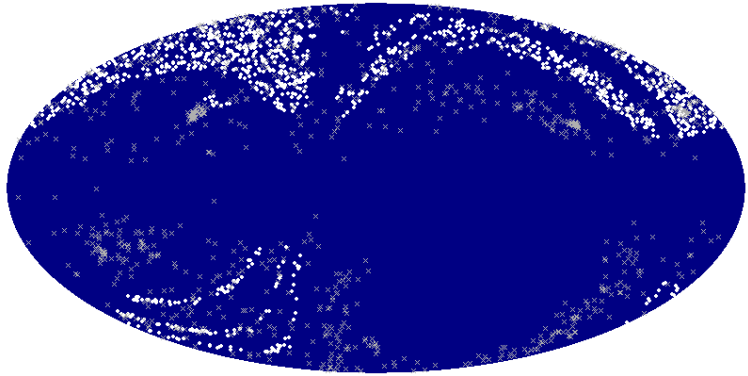,width=10cm}
}}
\caption{Sky positions of selected radio sources in Galactic
coordinates. The circles indicate SDSS objects and the crosses
mark all other sources.}
\label{f1:Verkhodanov2_n_en}
\end{figure*}

The improved quality of observations and parameter simulation
marks the approaching era of precision cosmology (as M.\,Longaire
named it in 2000 in Manchester). The recent space and ground-based
experiments \cite{wmap5ytem:Verkhodanov2_n_en,sdss:Verkhodanov2_n_en} have provided a wealth of data
for measuring the parameters of the Universe and constructing a
self-consistent cosmological mode. Nevertheless, checking whether
the $\Lambda$CDM paradigm is compatible with other tests remains a
priority task, as the current accuracy of the parameter
determination ($H_0$, $\Omega_\Lambda$, $\Omega_{DM}$, $\Omega_b$,
$\Omega_k$, etc.) still leaves the possibilities open for an
existence of other models~\cite{wmap5ycos:Verkhodanov2_n_en}. We place radio
galaxies among the most interesting objects for independent tests
of the above parameters. Radio galaxies are between the most
luminous galaxies known, and can hence be studied at large
redshifts, and thereby used to probe the state of the Universe in
other epochs. Of great importance for the study of radio galaxies
is the fact that their parent objects are giant elliptical
galaxies (gE), which can be used as standard candles/rulers
~\cite{sobol_dis:Verkhodanov2_n_en} at the initial stage of selection.
Identification with gE is important both for tracing the evolution
of stellar systems at large redshifts, and searching for distant
groups of galaxies or protoclusters in centers of which they
reside, as well as studying mergers and interactions, which can be
singled out by the manifestations of galactic nuclear activity.
Parameters, determined from observational data, include the age of
the \mbox{galaxy \cite{rg_age:Verkhodanov2_n_en,rg_rc_age:Verkhodanov2_n_en,age_sys:Verkhodanov2_n_en},} which is
limited by the age of the Universe---it cannot exceed the latter,
since galaxies need certain time to form after the birth of the
Universe. The age of the Universe, in its turn, is determined by
the Hubble parameter. One of the specific features of radio
galaxies that makes them convenient objects for chronometric
estimates is an almost passive stellar evolution of their parent
galaxies  even at large redshifts  ($z\sim4$)
\cite{willott_k_z:Verkhodanov2_n_en,verkh_phot_z:Verkhodanov2_n_en}, which allows estimating the ages
of the systems.

When constructed, a complete distribution of radio galaxies on
their redshifts $z$ can be used not only to study the luminosity
function of these objects, but to investigate the formation of
supermassive black holes at the galactic centers, and the dynamics
of the expansion of the Universe. According to the observational
data of Disney et al.~\cite{rg6pars:Verkhodanov2_n_en}, the physical parameters of
radio galaxies (total mass, baryon fraction, age, luminosity,
etc.) are intercorrelated, and this fact should make it easier to
use gE type galaxies in cosmological studies, provided that all
these objects formed at the same epoch. However, to address such
tasks, one needs the most complete sample of radio galaxies
possible in different redshift intervals.

The aim of this paper is to construct a sample of radio galaxies
in the redshift interval of \mbox{$z>0.3$.} It is the second part
of the catalog published after the Paper I~\cite{rg_listI:Verkhodanov2_n_en}, and
contains the catalogued optical data for radio galaxies adopted
from other available sources.

\begin{figure*}[tbp]
\vspace{0.2cm}
\centerline{
\vbox{
\hbox{
\psfig{figure=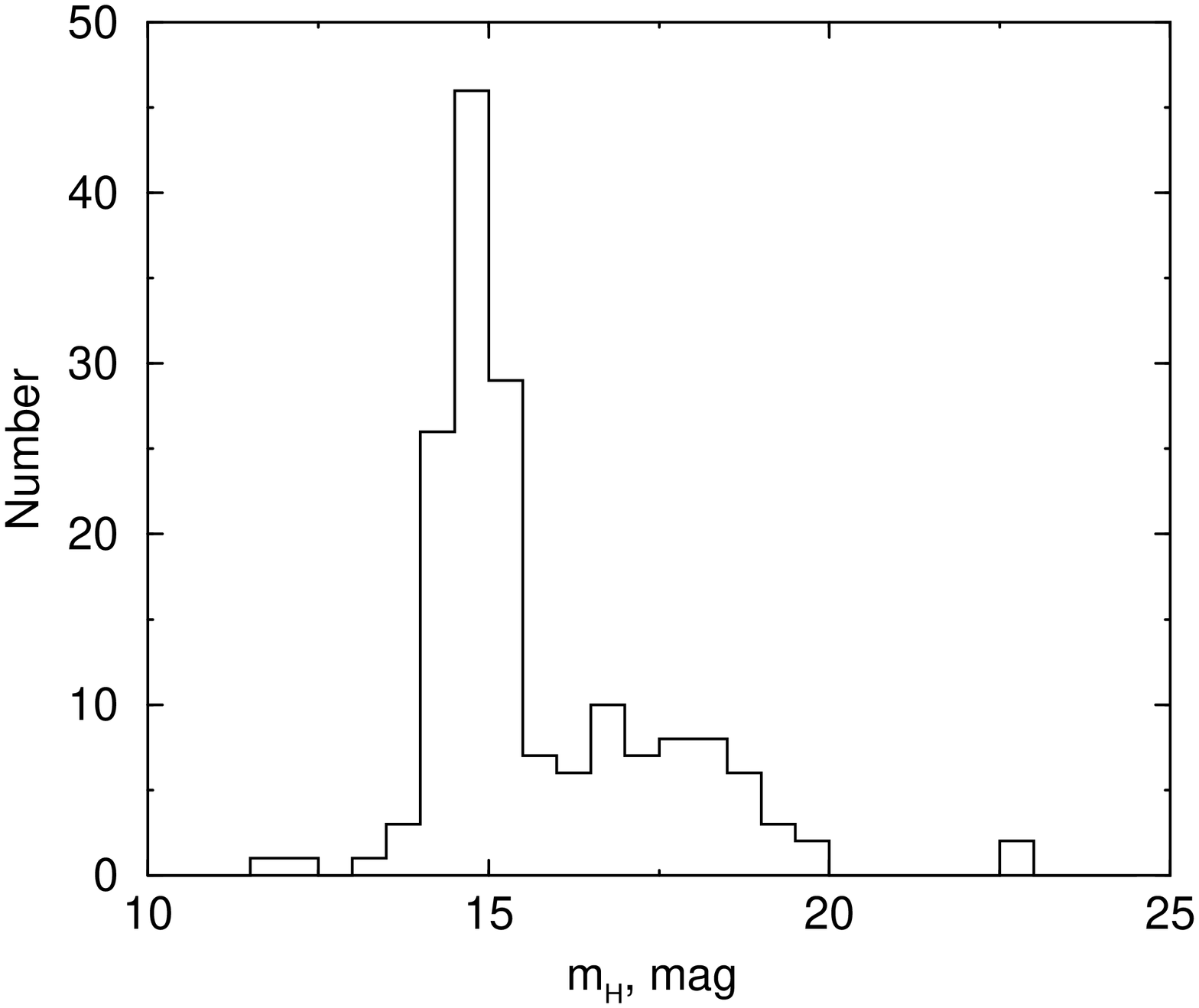,width=5.3cm,angle=-90}
\psfig{figure=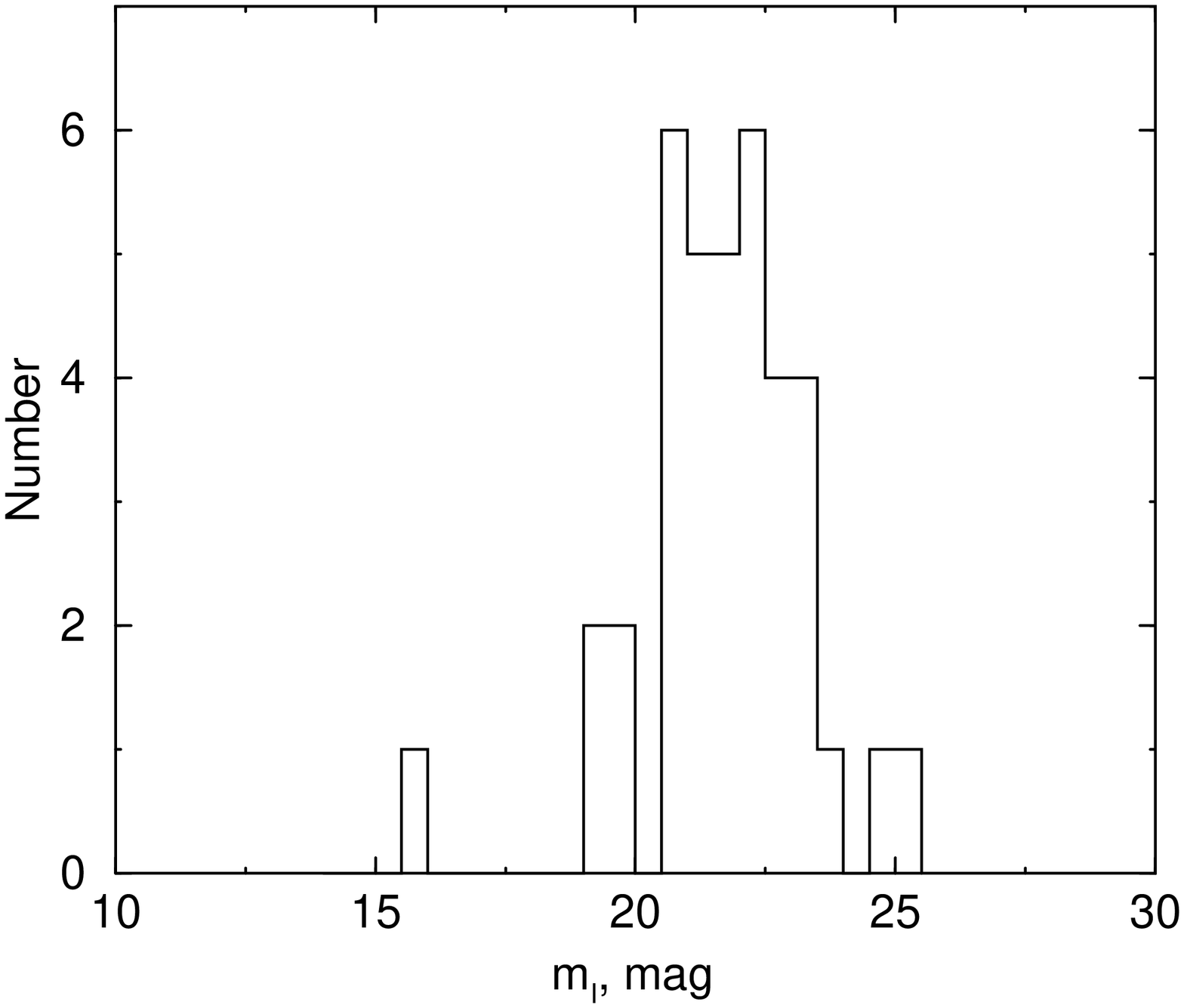,width=5.3cm,angle=-90}
\psfig{figure=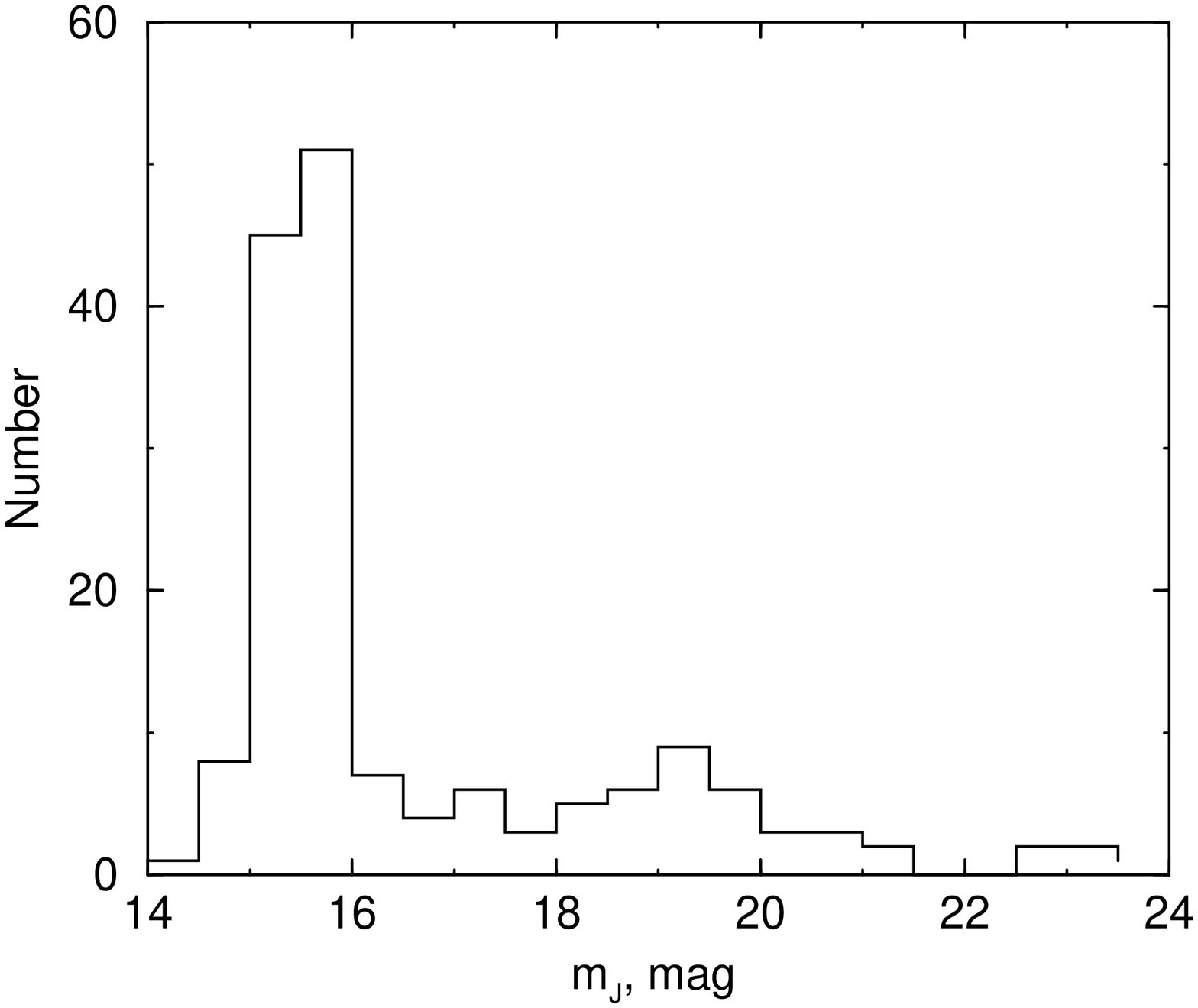,width=5.3cm,angle=-90}
}
\hbox{
\psfig{figure=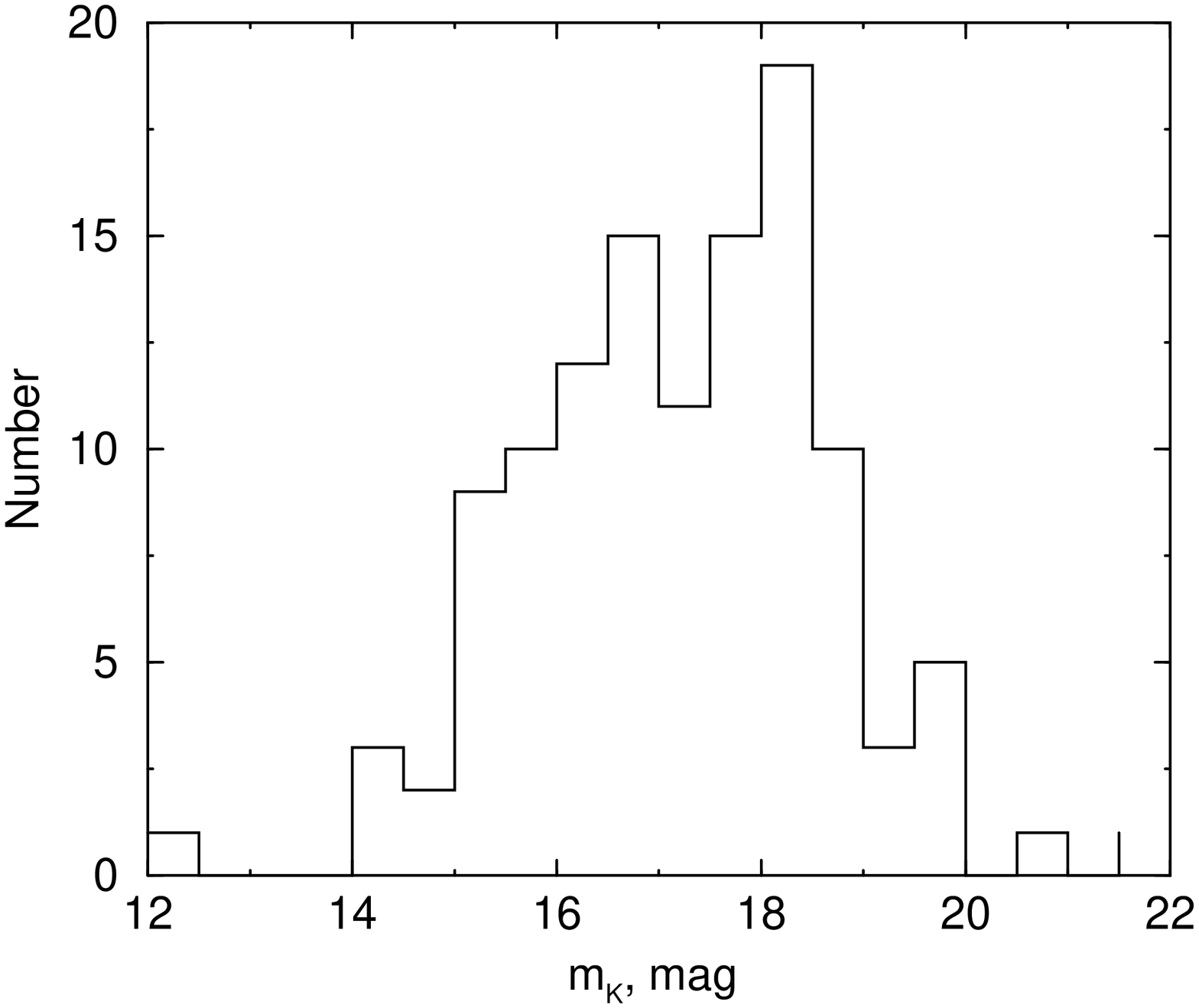,width=5.3cm,angle=-90}
\psfig{figure=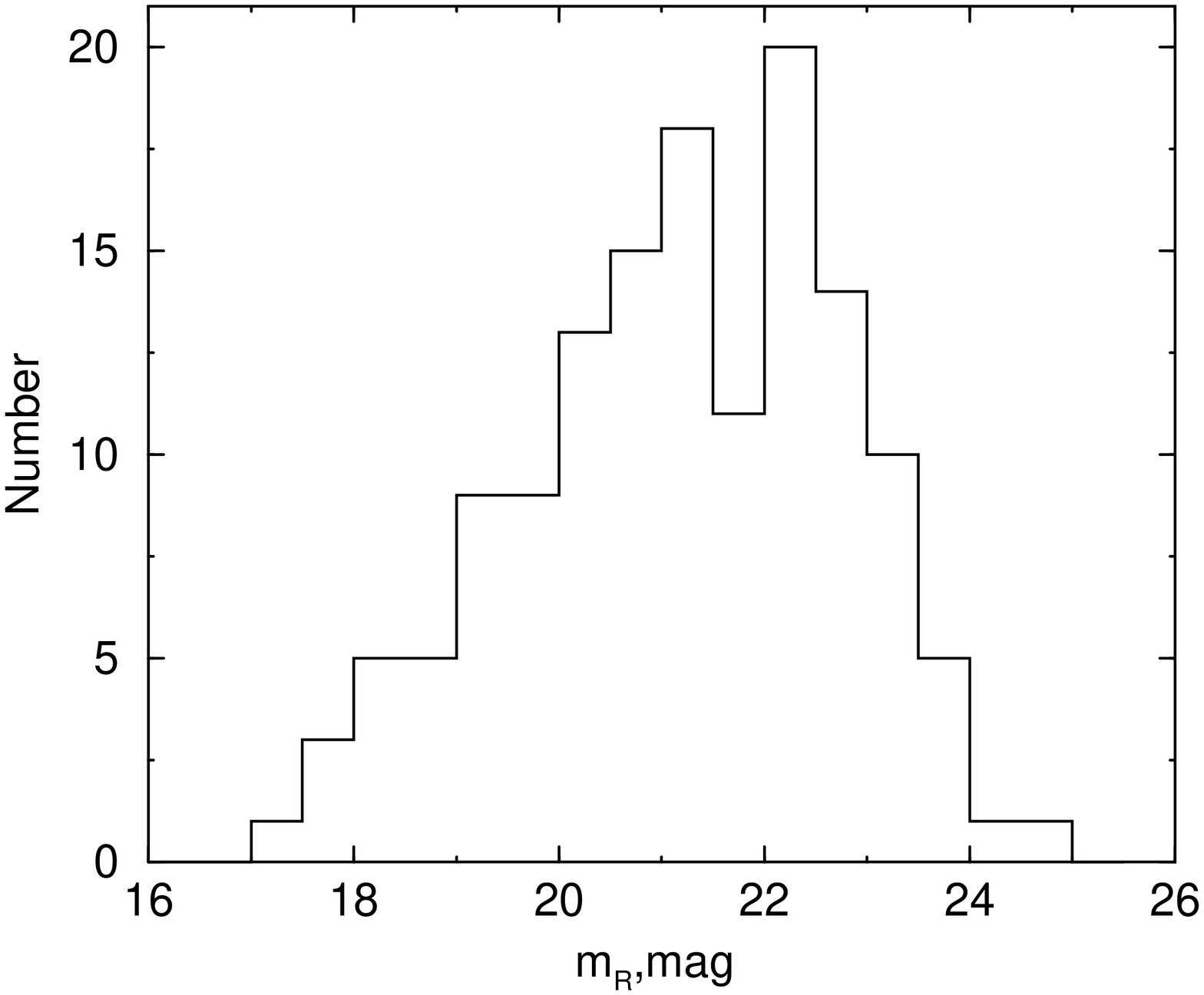,width=5.3cm,angle=-90}
\psfig{figure=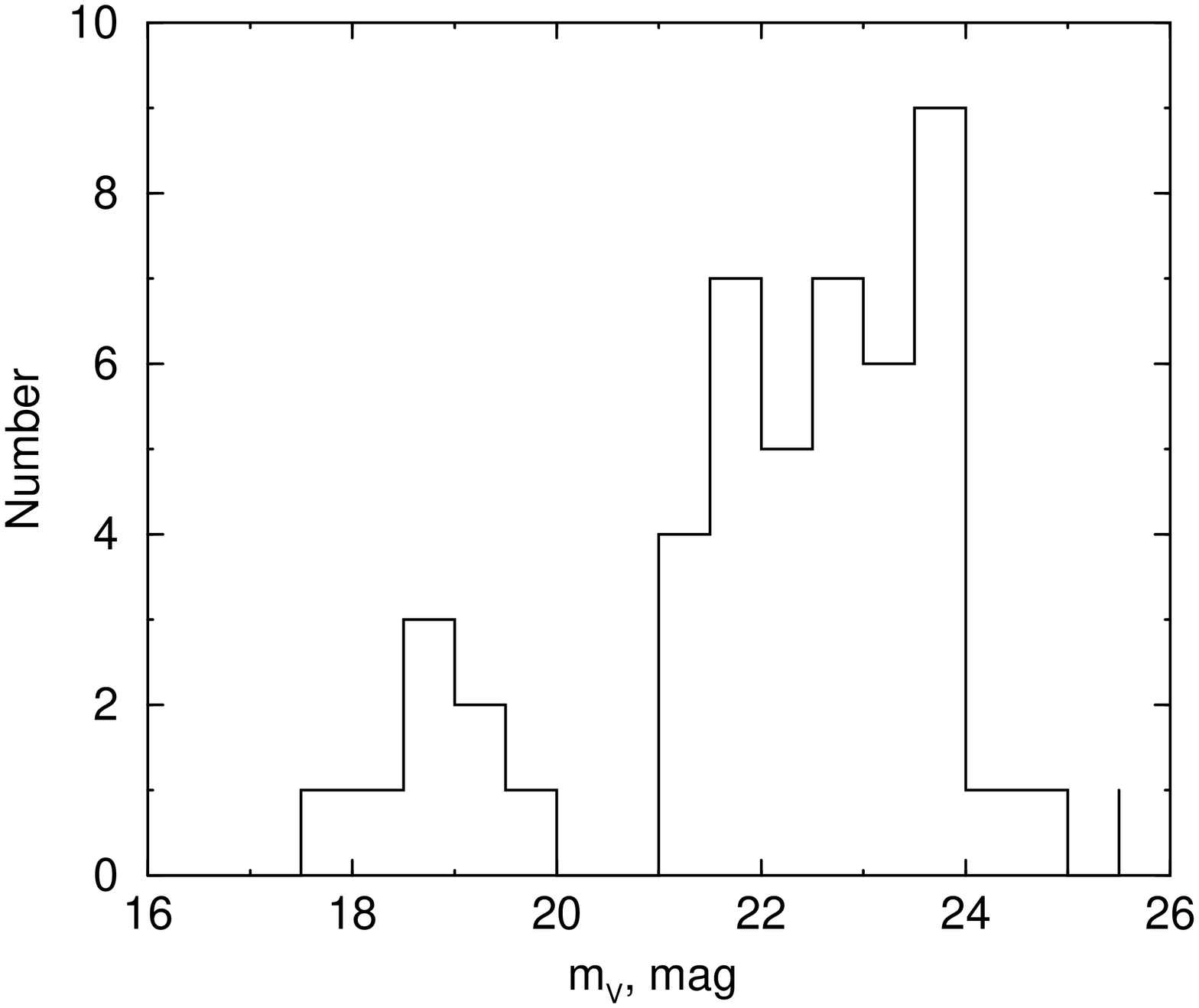,width=5.3cm,angle=-90}
}
\hbox{
\psfig{figure=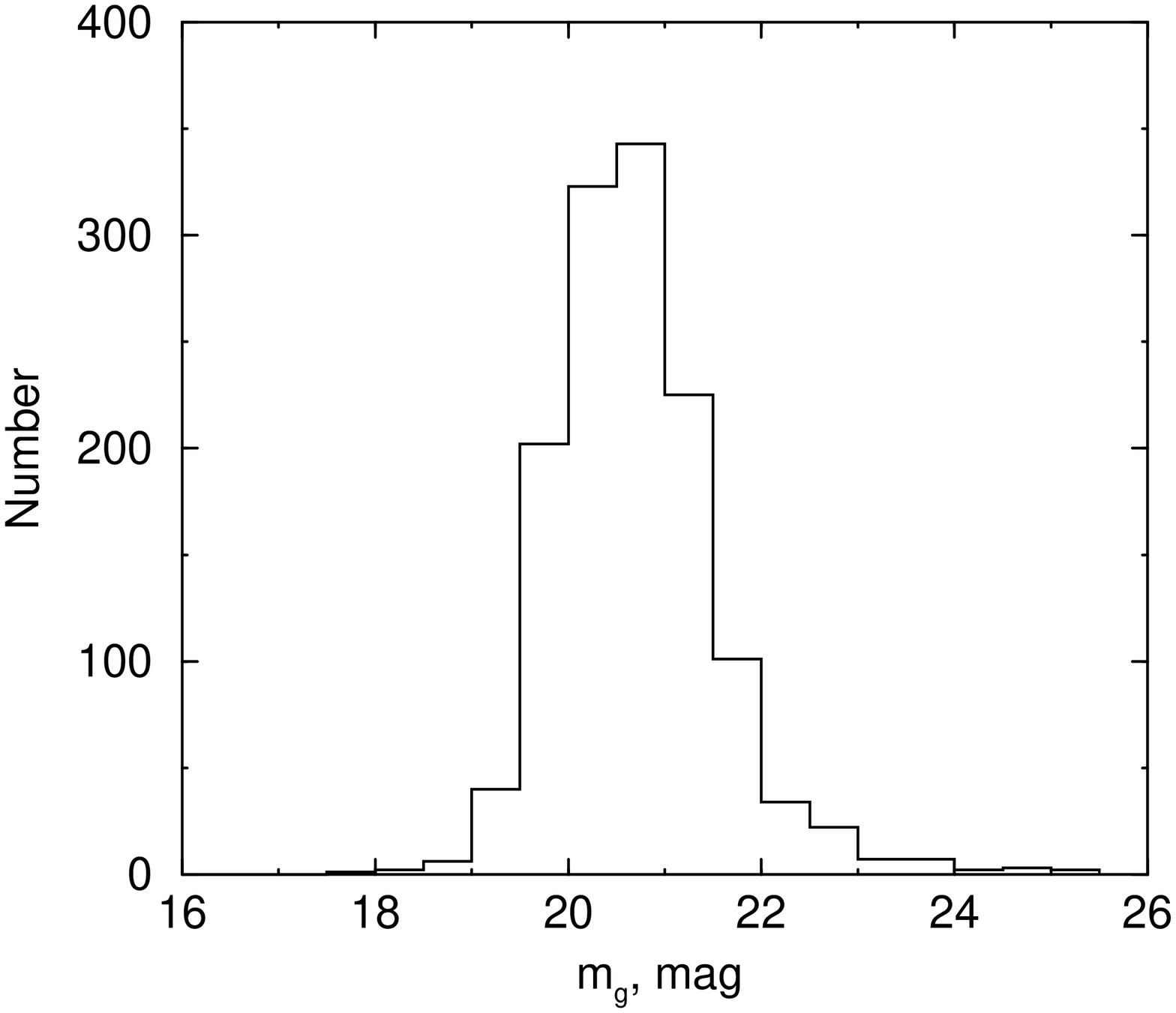,width=5.3cm,angle=-90}
\psfig{figure=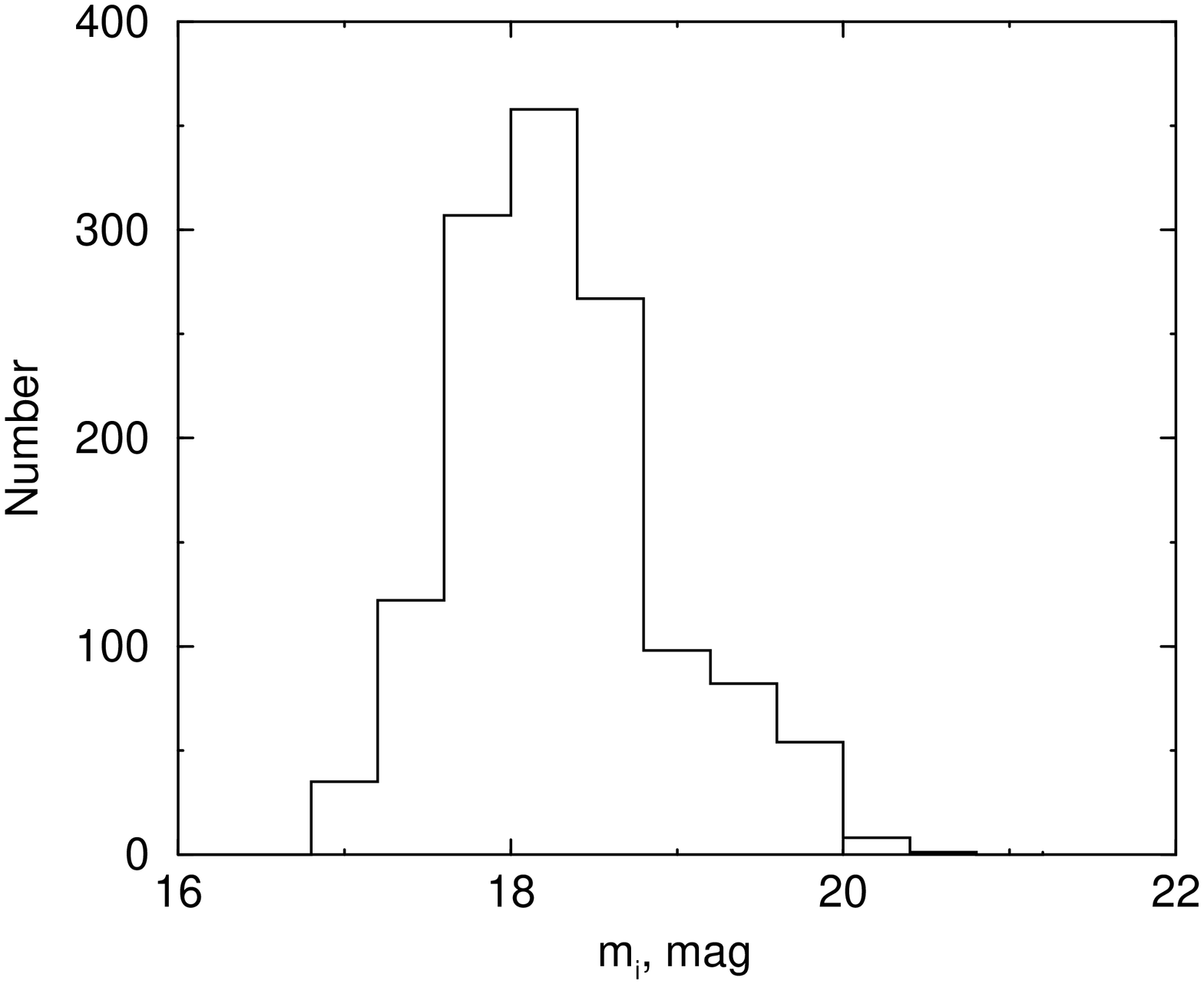,width=5.3cm,angle=-90}
\psfig{figure=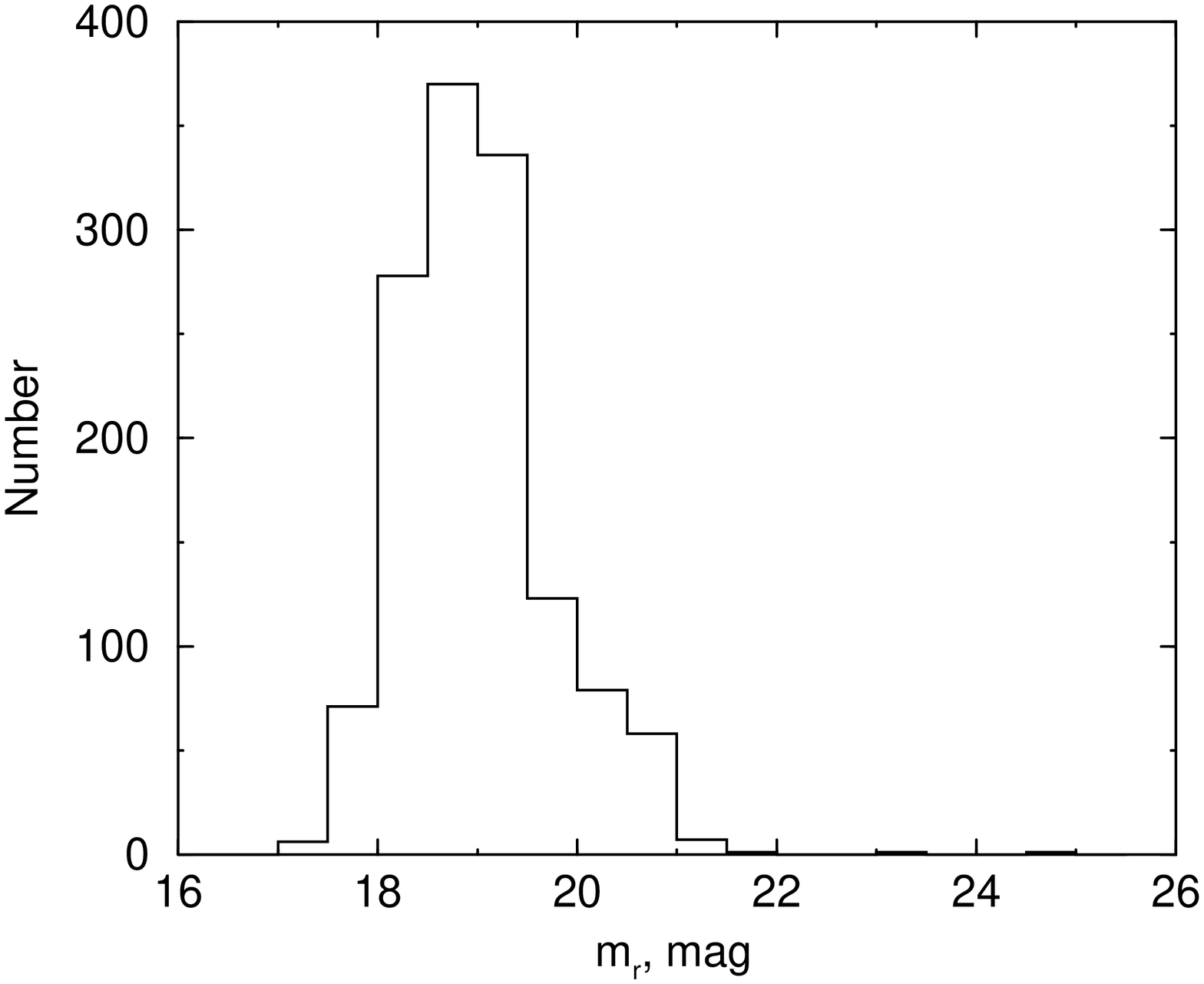,width=5.3cm,angle=-90}
}
\hbox{
\psfig{figure=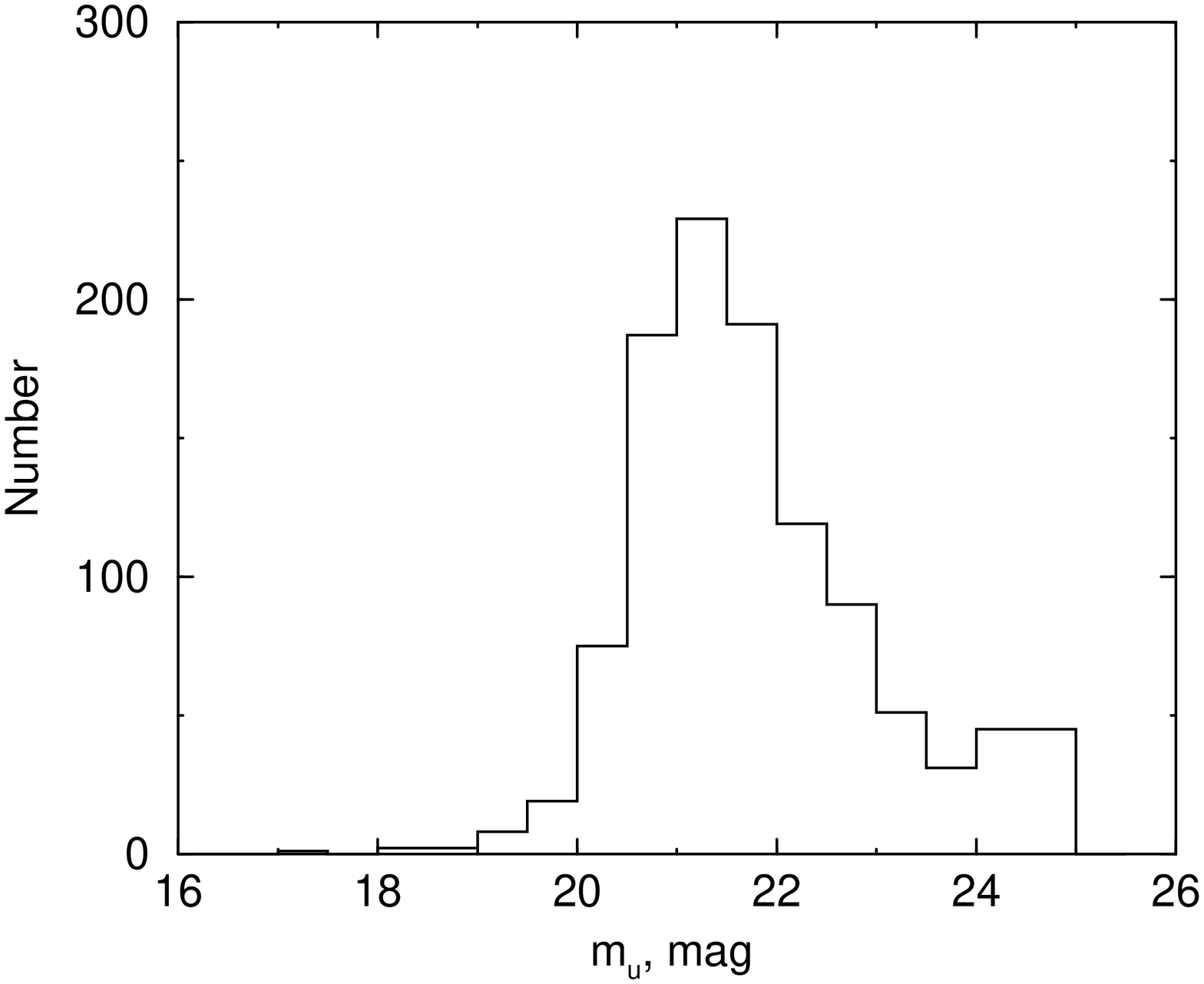,width=5.3cm,angle=-90}
\psfig{figure=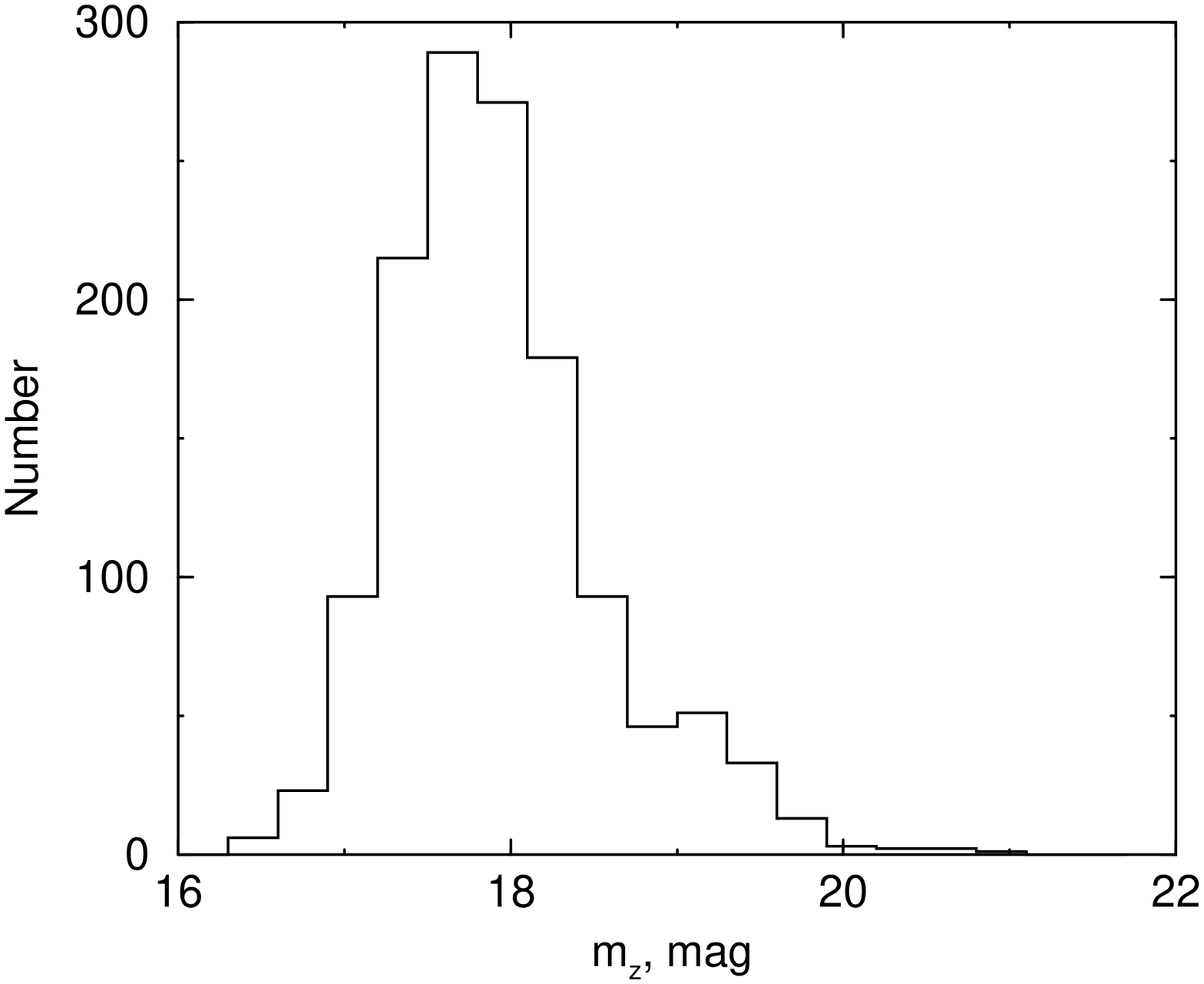,width=5.3cm,angle=-90}
}
}}
\caption{Left to right, top to bottom: histograms of the  H, I, J, K, R, V, g, r, u,
i, and z-band magnitudes.}
\label{f2:Verkhodanov2_n_en}
\end{figure*}

\begin{figure*}[tbp]
\centerline{
\vbox{
\hbox{
\psfig{figure=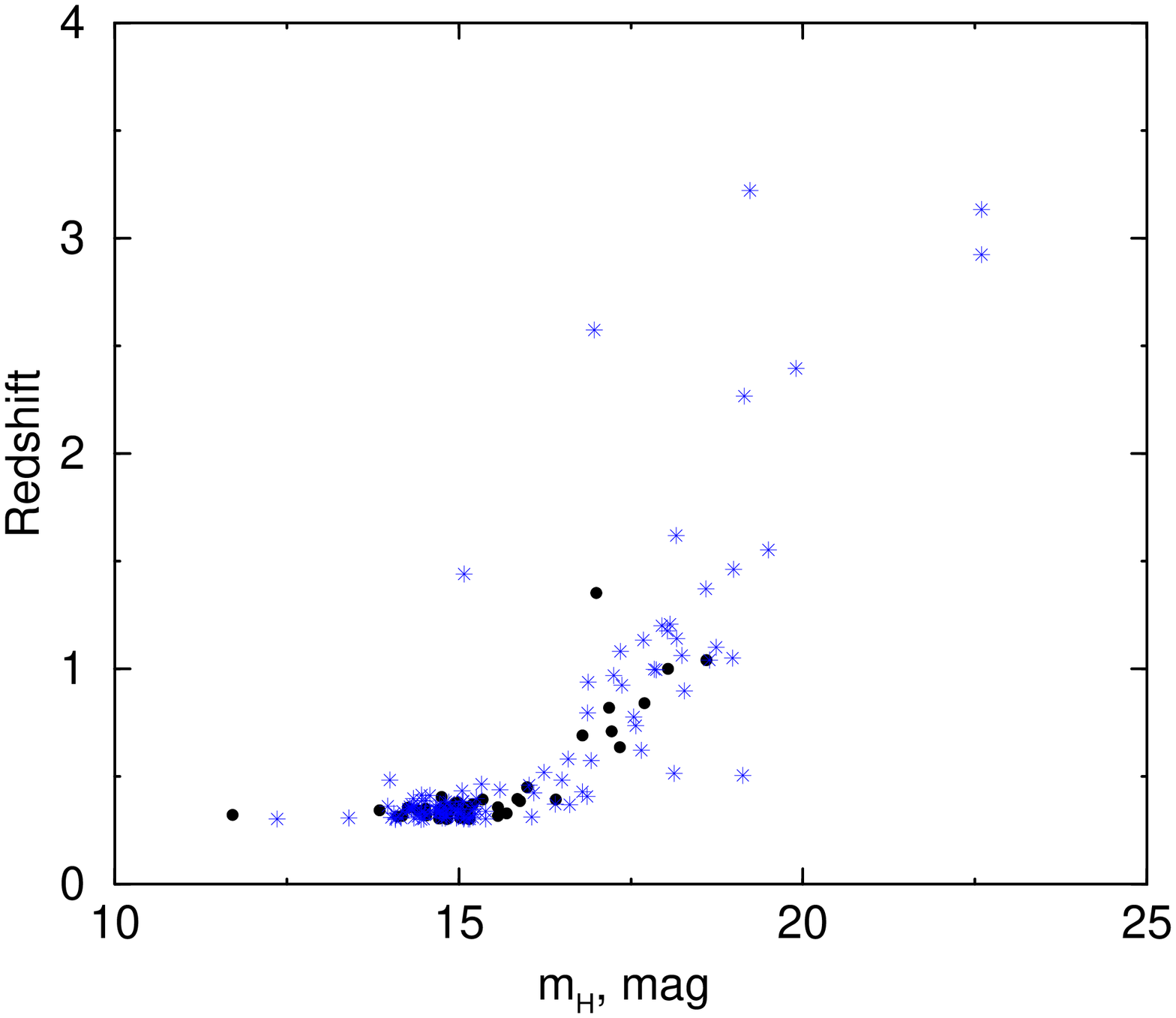,width=8cm,angle=-90}
\psfig{figure=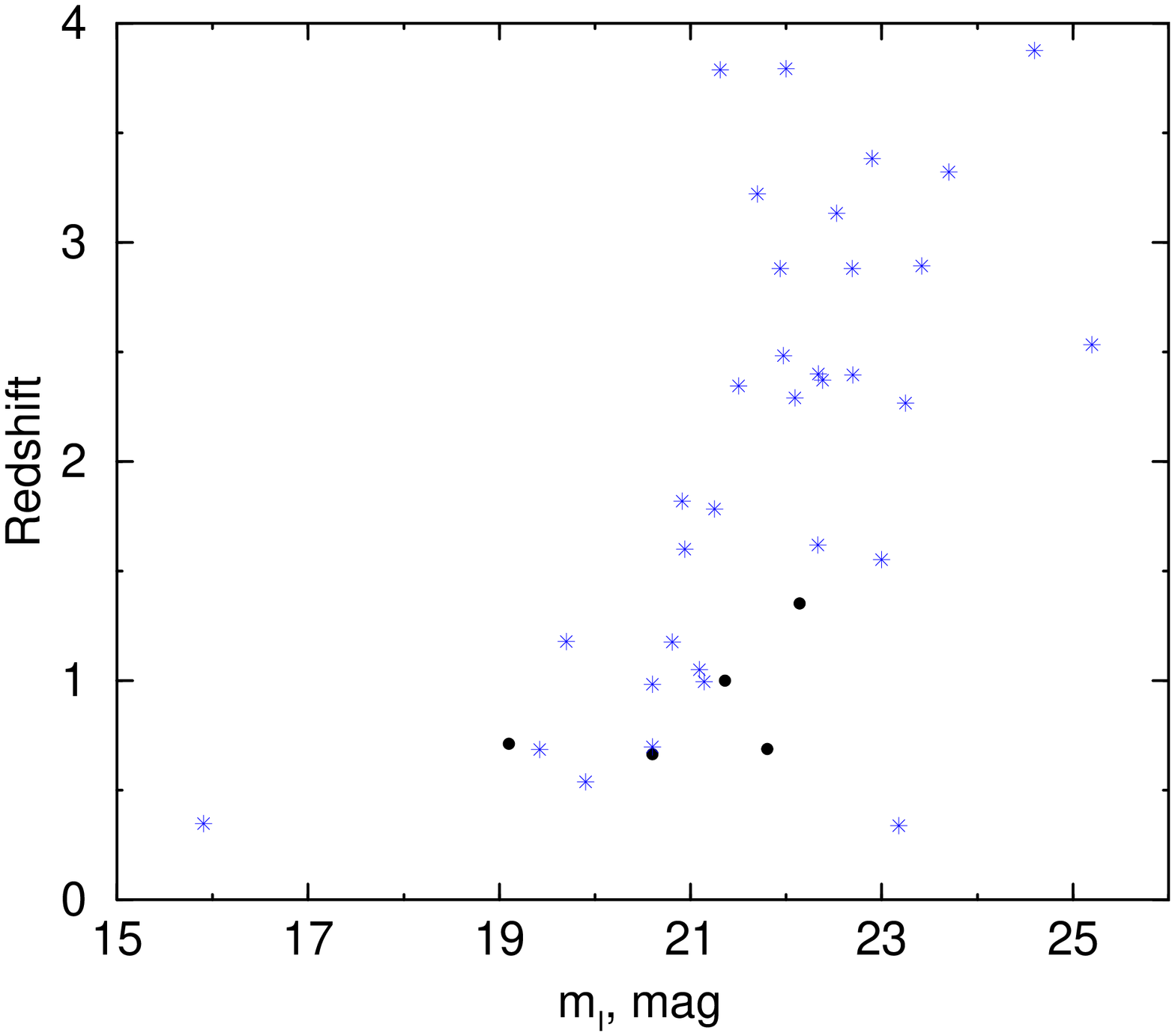,width=8cm,angle=-90}
}
\hbox{
\psfig{figure=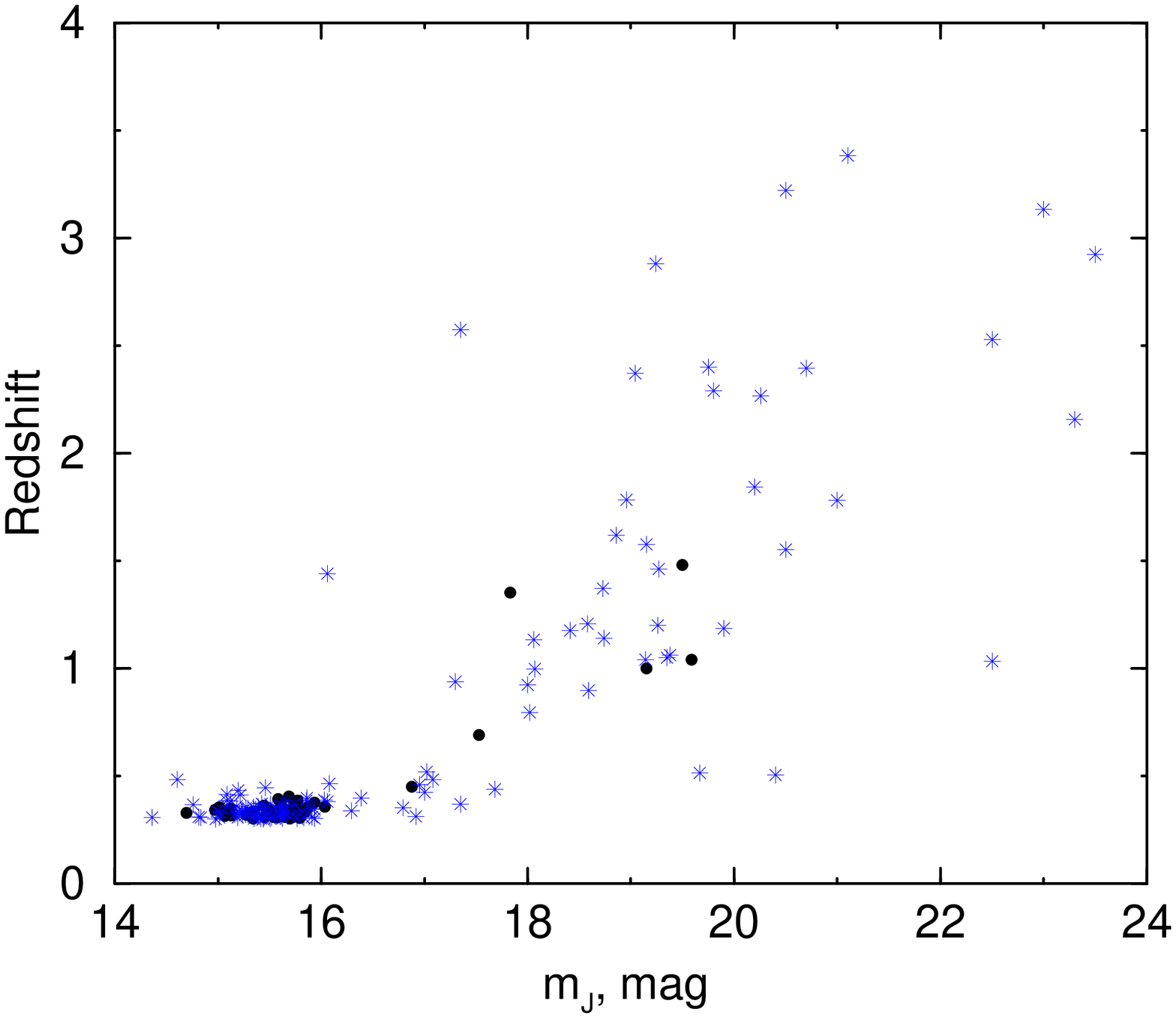,width=8cm,angle=-90}
\psfig{figure=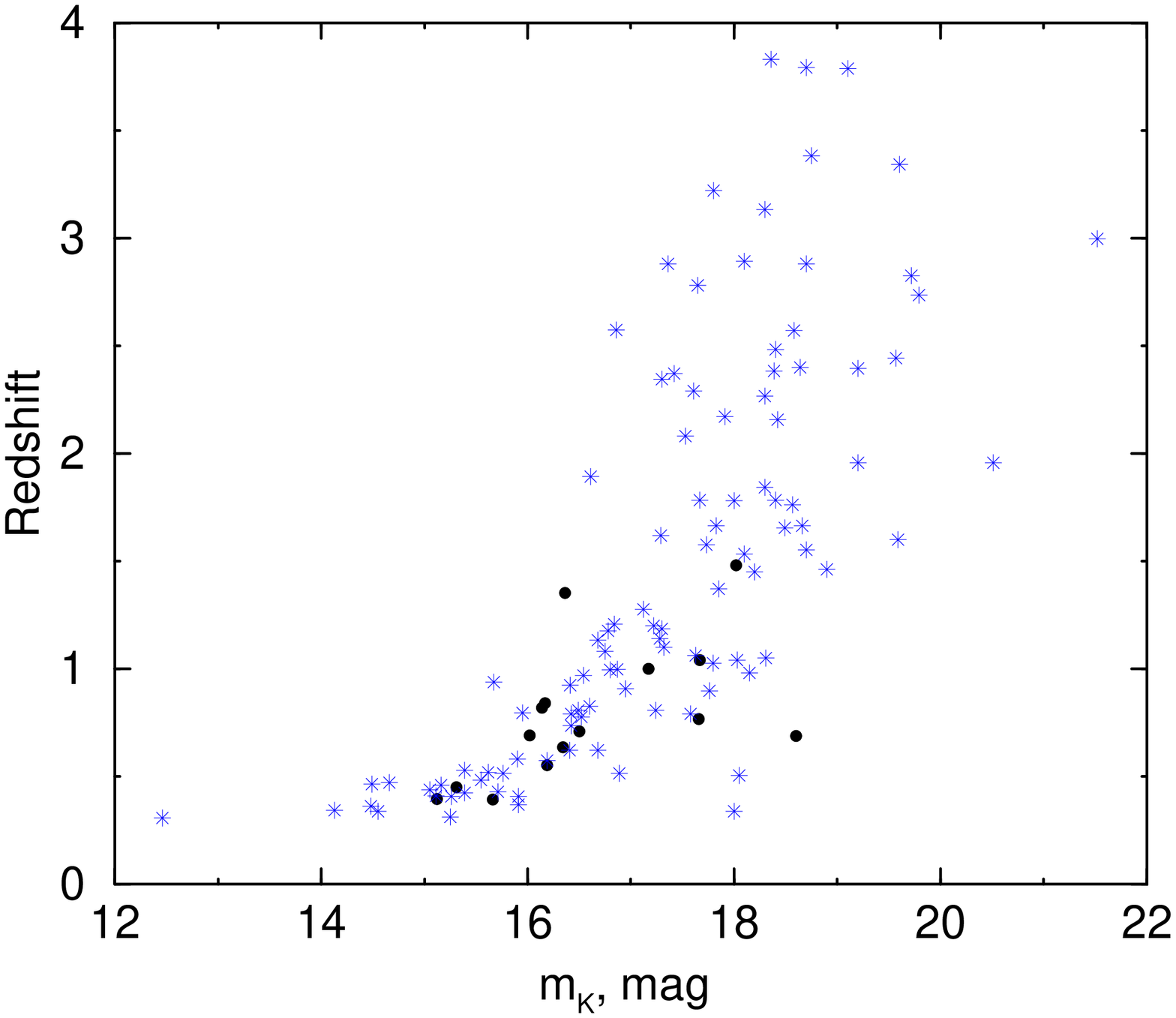,width=8cm,angle=-90}
}
\hbox{
\psfig{figure=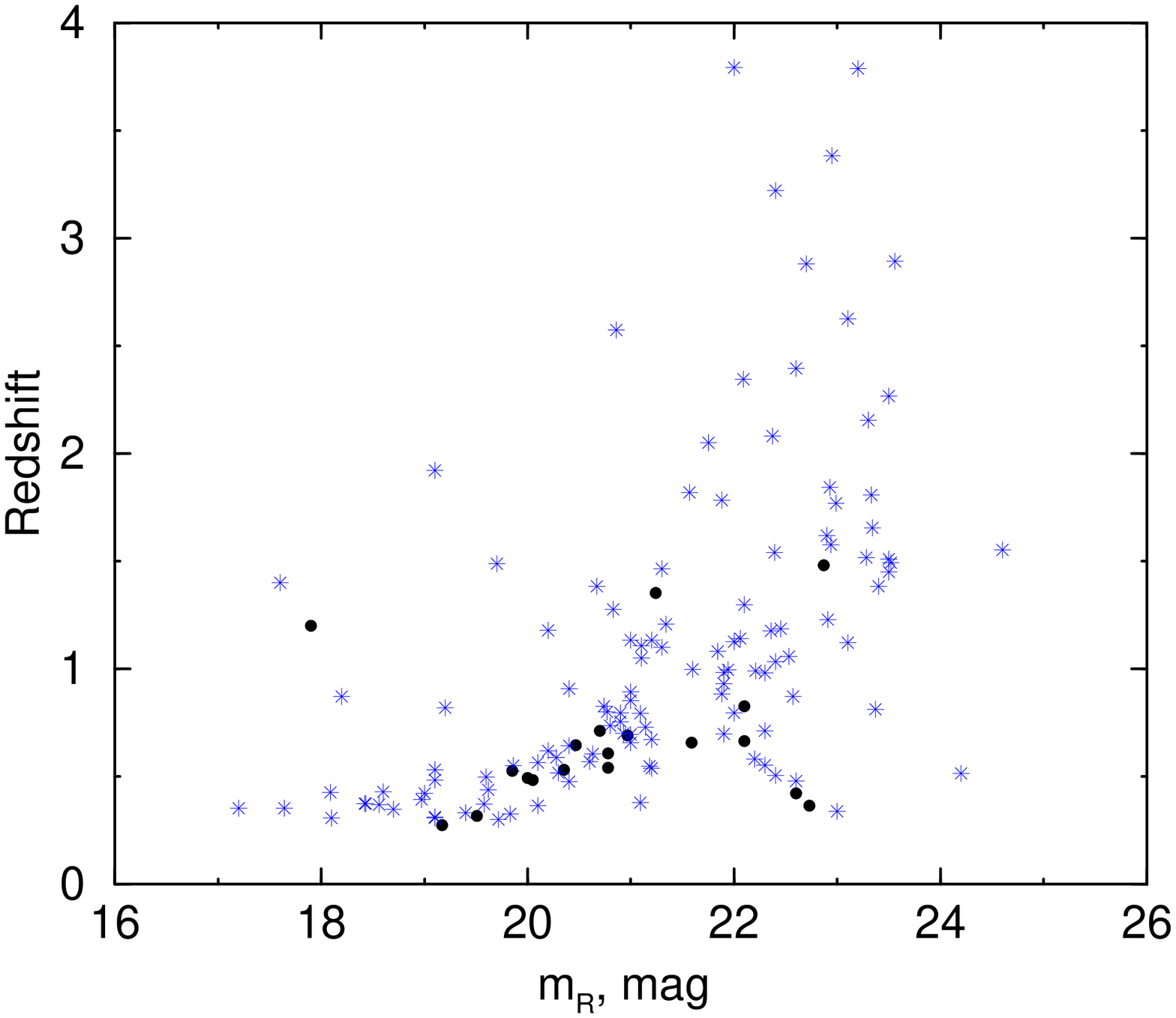,width=8cm,angle=-90}
\psfig{figure=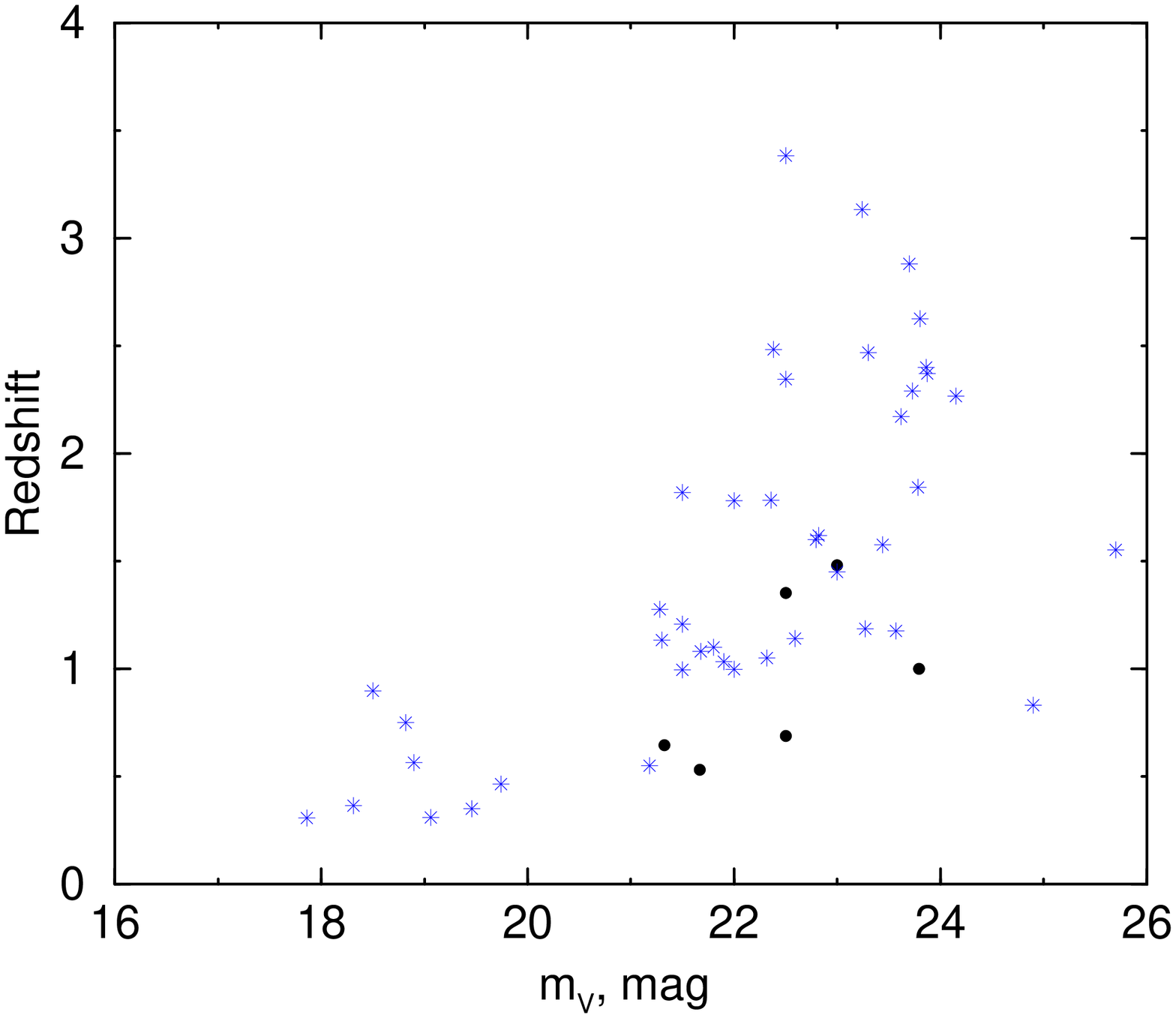,width=8cm,angle=-90}
} }}
\caption{Left to right, top to bottom: the ``magnitude--$z$''
Hubble diagrams for the  H, I, J, K, R, and V passbands. The
galaxies from groups (1) and (2) are marked with crosses and}
\end{figure*}

\begin{figure*}[tbp]
\vspace{0.3cm}
\centerline{
\vbox{
\hbox{
\psfig{figure=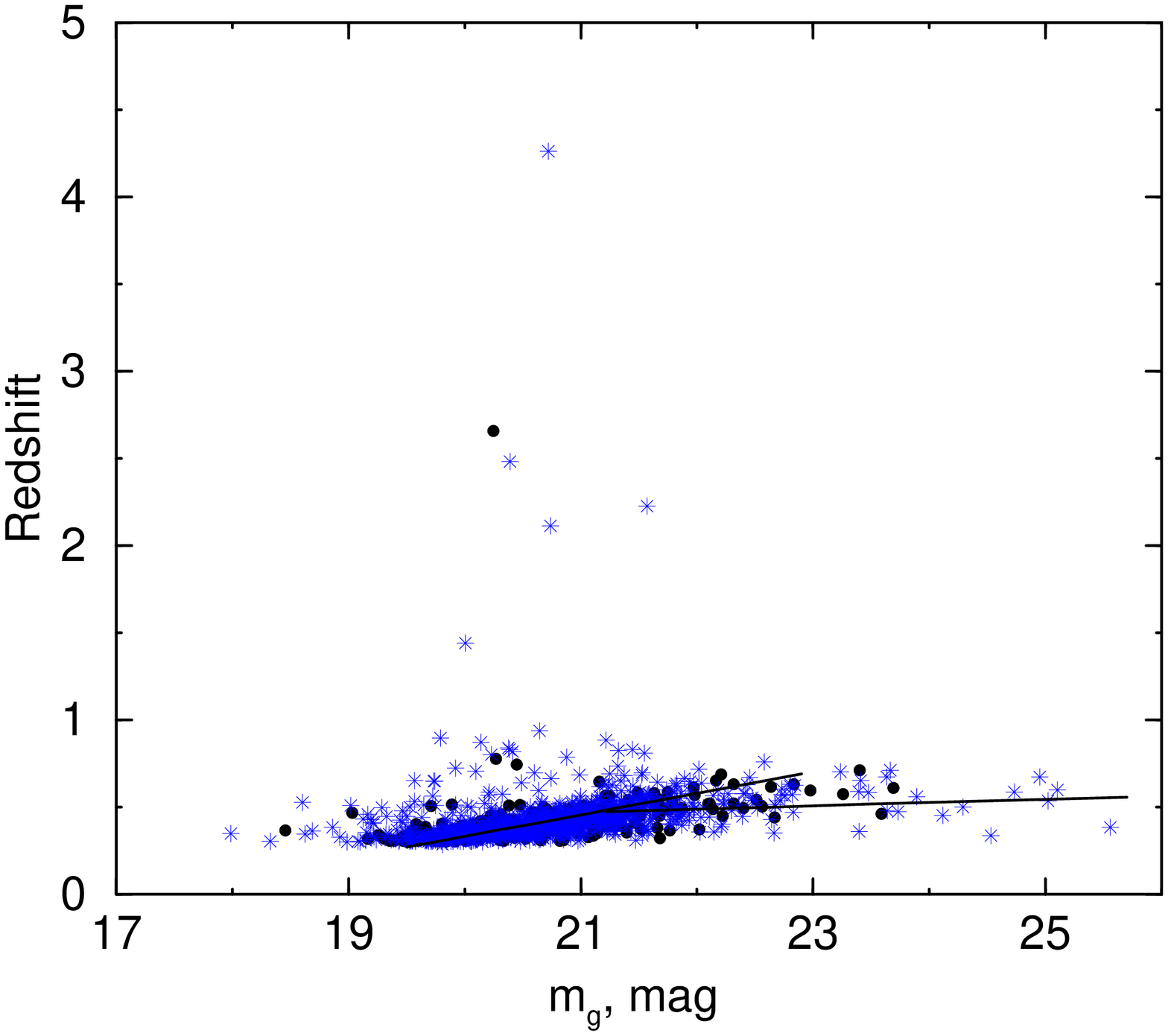,width=7.5cm,angle=-90}
\psfig{figure=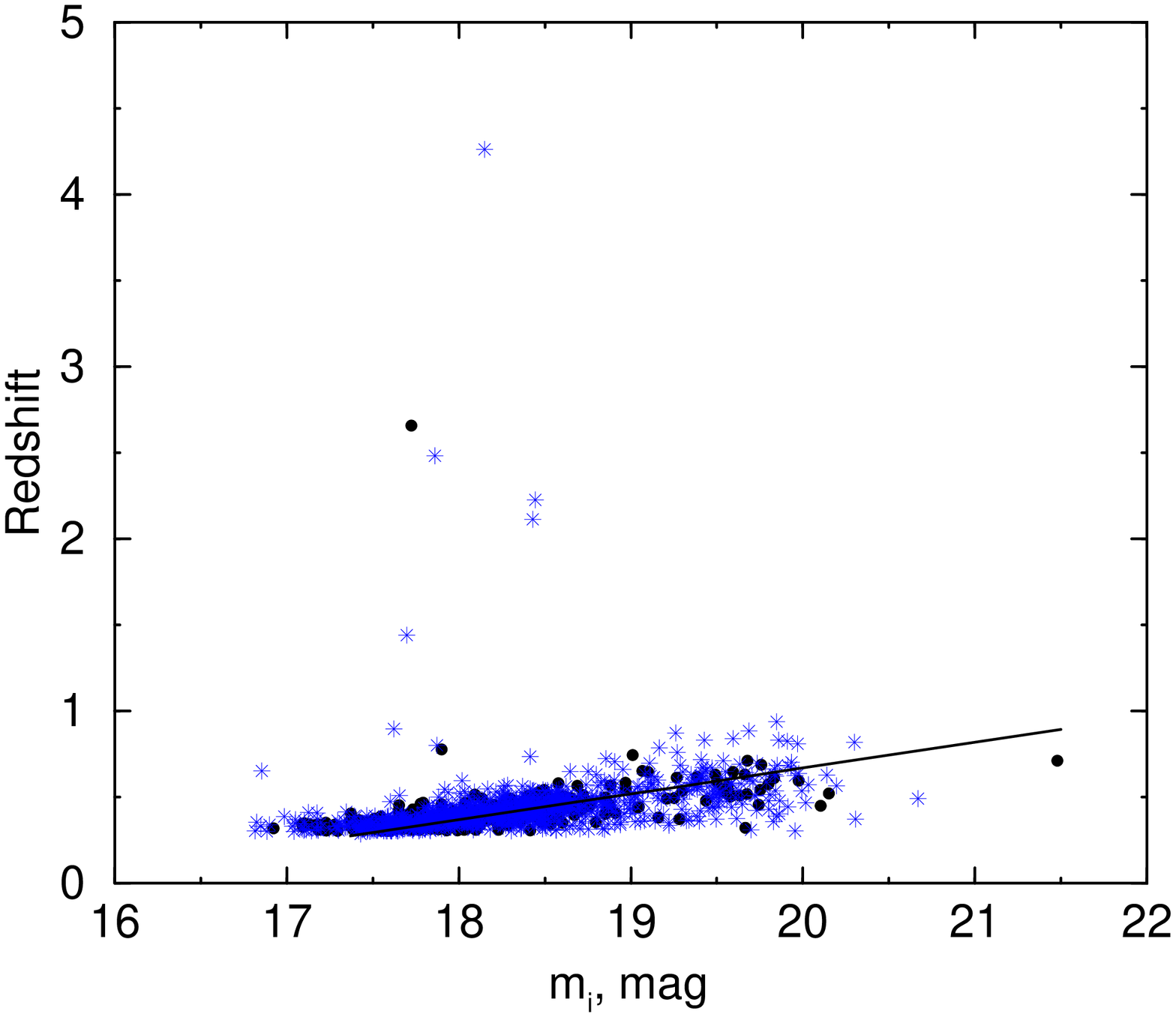,width=7.5cm,angle=-90}
}
\hbox{
\psfig{figure=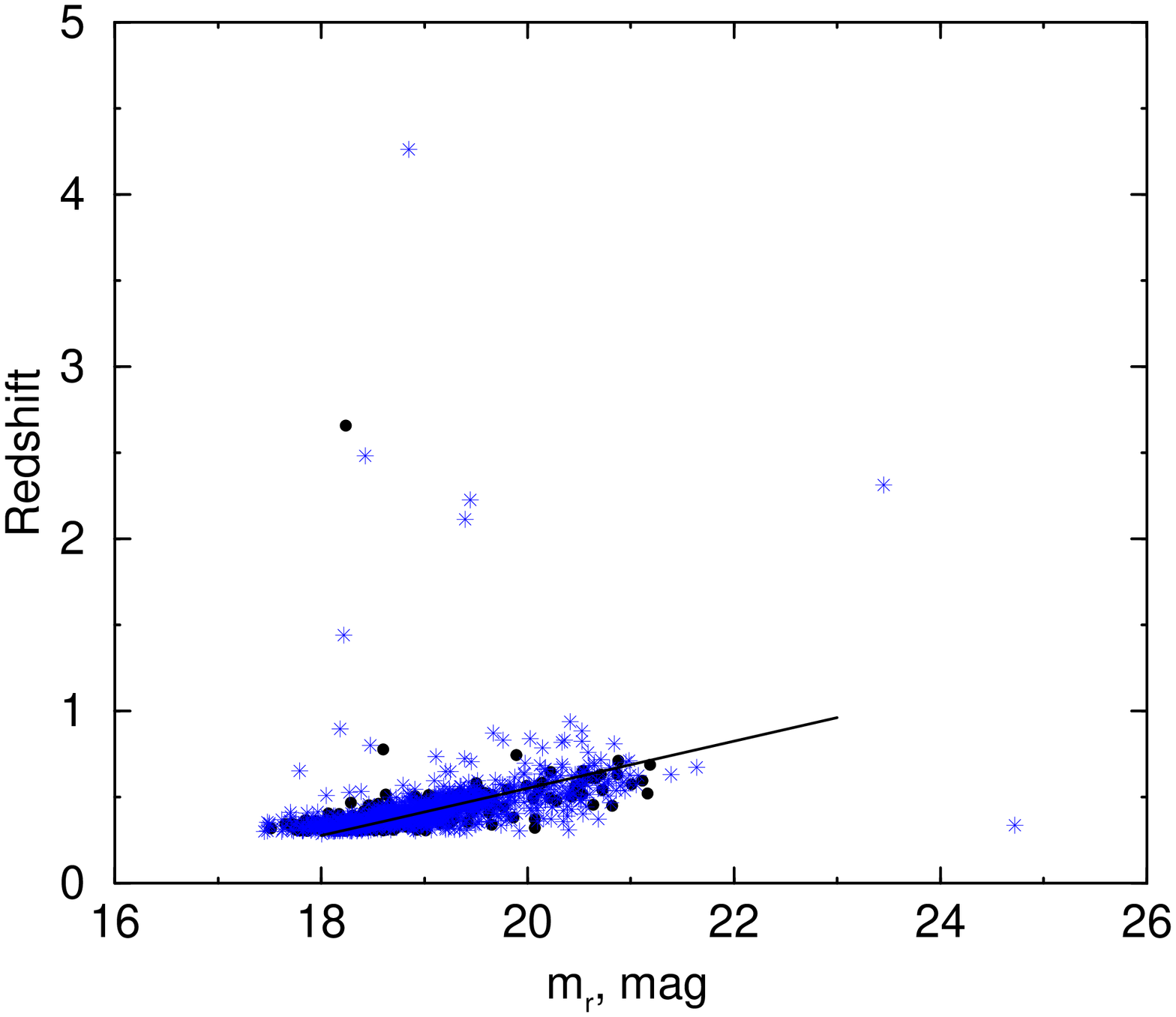,width=7.5cm,angle=-90}
\psfig{figure=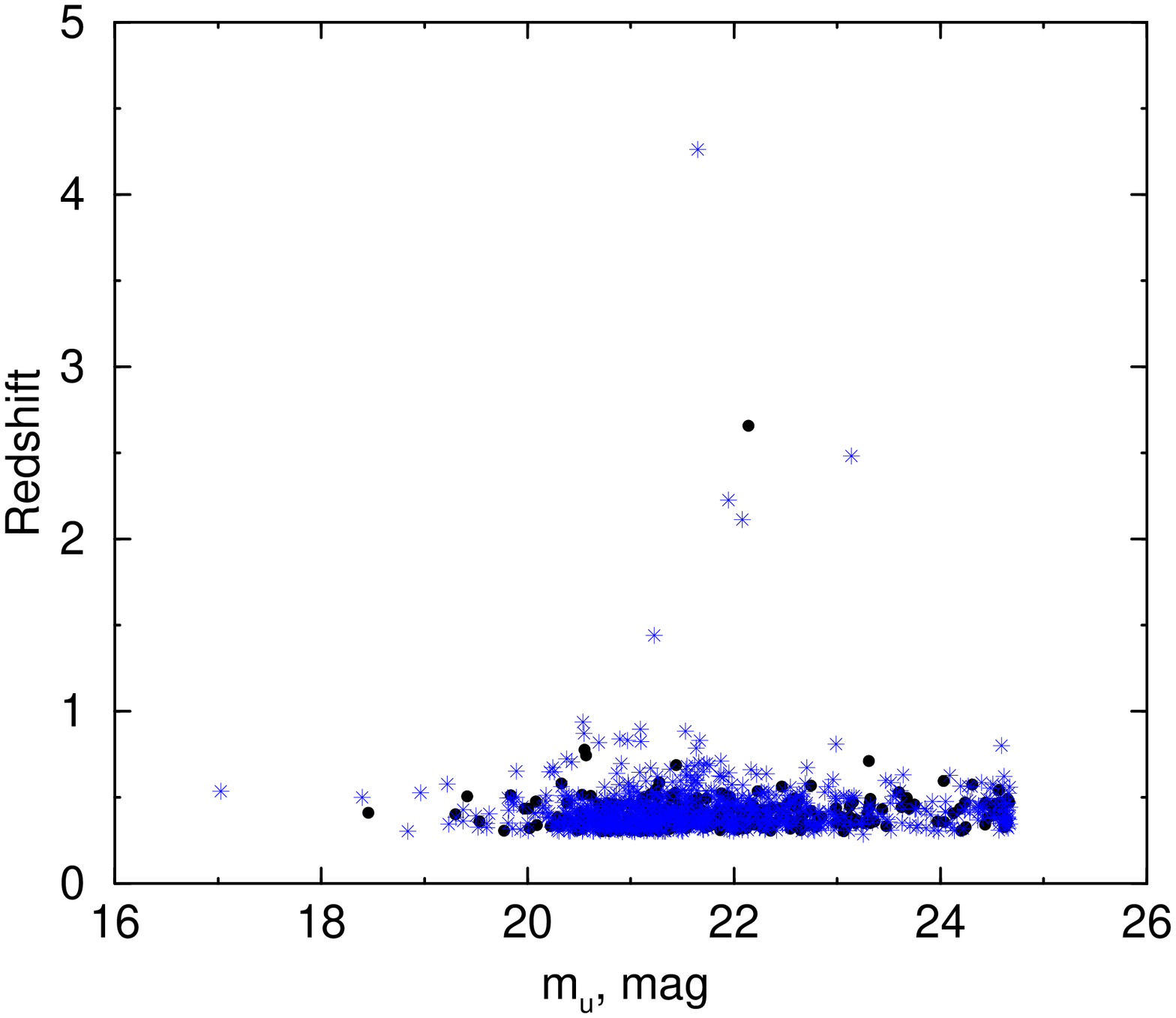,width=7.5cm,angle=-90}
}
\hbox{
\psfig{figure=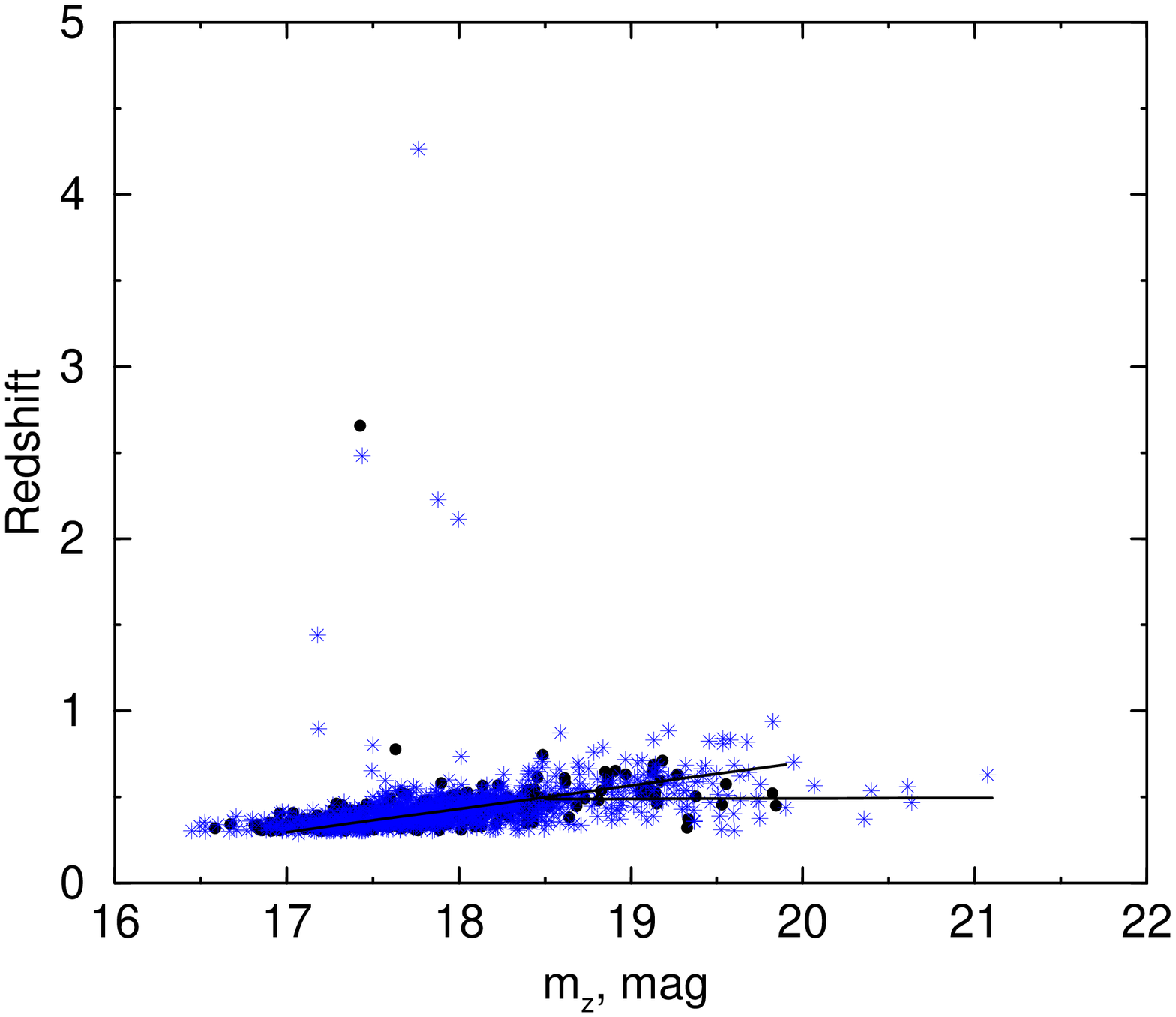,width=7.5cm,angle=-90}
\mbox{\hspace*{7.5cm}} } }}
\caption{Left to right, top to bottom:
the ``magnitude--$z$'' Hubble diagrams for the g, i, r, u, and z
passbands. The galaxies from groups (1) and (2) are marked with
crosses and circles, respectively. Regression relations $mag(z)$
are derived for the g, i, r, and z-band magnitudes. }
\label{f4:Verkhodanov2_n_en}
\end{figure*}

\begin{figure*}[tbp]
\vspace{0.5cm}
\centerline{
\hbox{
\psfig{figure=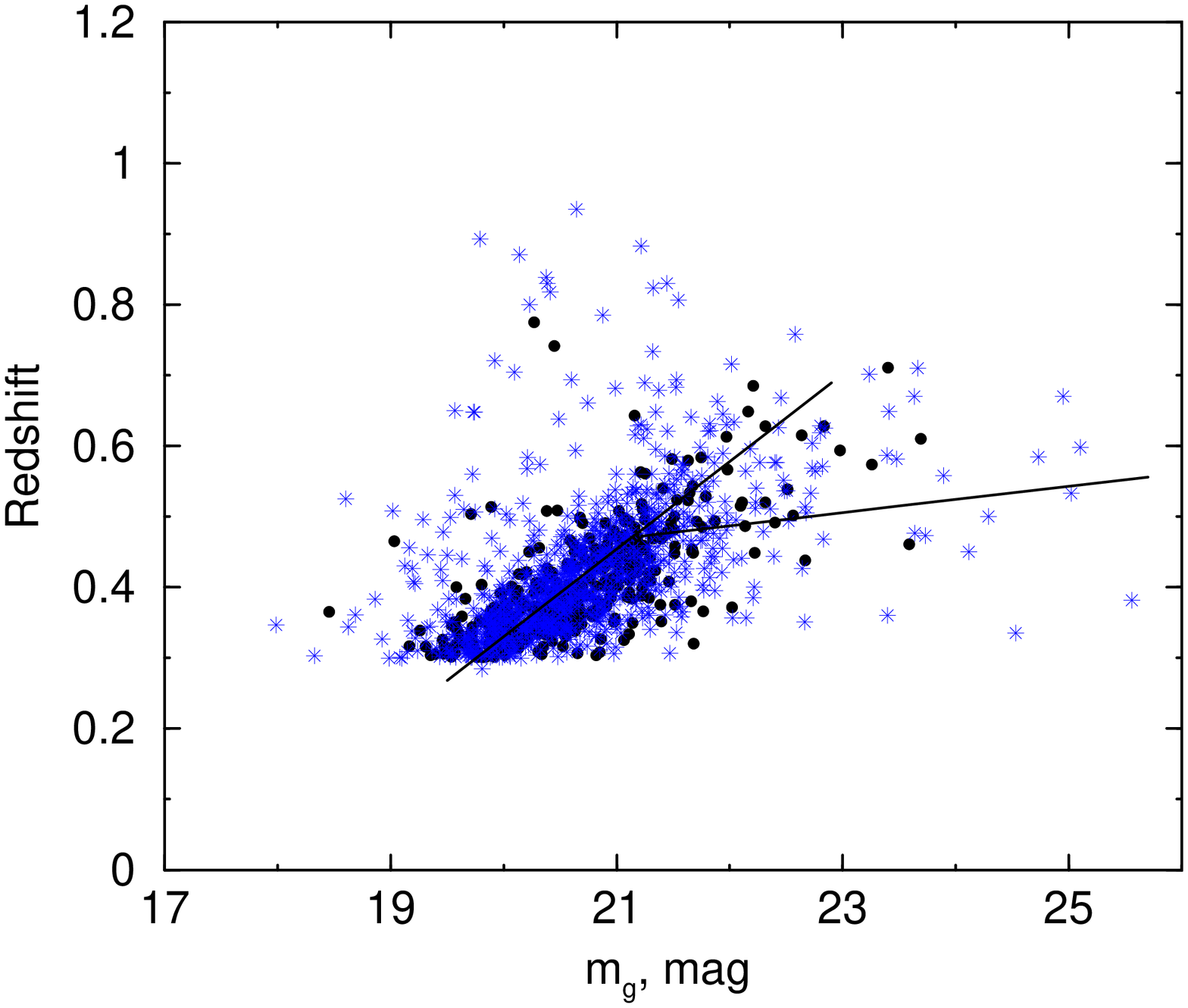,width=8cm,angle=-90}
\psfig{figure=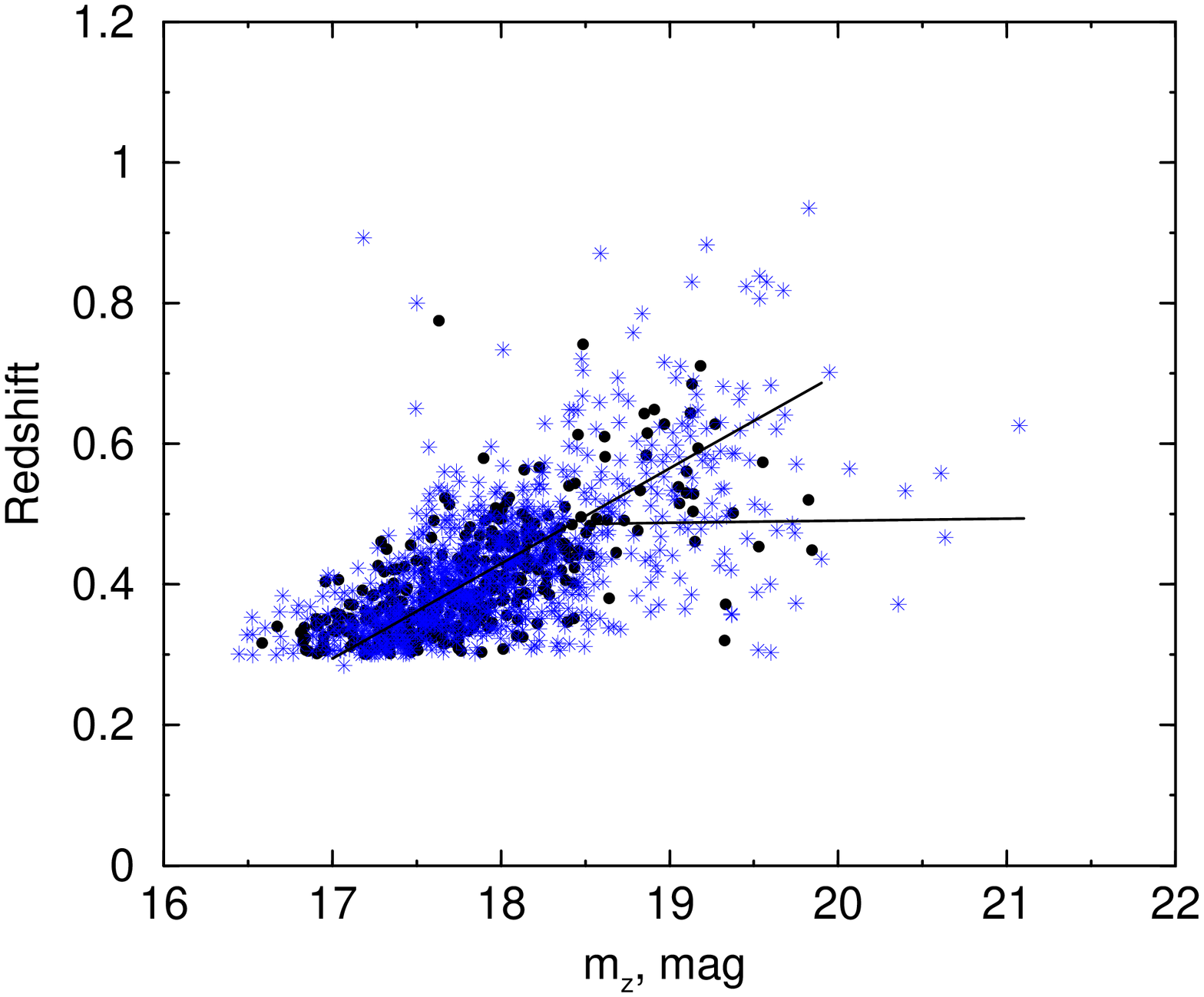,width=8cm,angle=-90}
}}
\caption{ The Hubble diagrams for the g- (left) and z-band
(right) filters. The galaxies from groups (1) and (2) are marked
with crosses and circles, respectively. The regression lines
emphasize the two object populations identified. }
\vspace{0.5cm}
\end{figure*}

Paper~I describes the construction of the sample using the
NED\footnote{\tt http://nedwww.ipac.caltech.edu} and
CATS\footnote{\tt http://cats.sao.ru} databases, and applying
selection procedures, that results in singling out a total of 2442
objects. The list of objects in Paper~I also contains the names of
the initial catalogs from where we adopted the data for the
objects (and the names of radio sources in case of the 3C and 4C
surveys), equatorial coordinates (J2000.0), and spectroscopic
redshifts. Note that the catalog of selected galaxies also
contains objects from other lists, including the sources with
ultrasteep spectra~\mbox{($\alpha<-1.0$, $S\sim\nu^\alpha$).}
These very objects are characterized by a high percentage of
distant radio galaxies
\cite{dagkes:Verkhodanov2_n_en,blum_miley:Verkhodanov2_n_en,par_big3a:Verkhodanov2_n_en,par_big3b:Verkhodanov2_n_en,uss_list:Verkhodanov2_n_en,miley_debreuck:Verkhodanov2_n_en}.
The most distant radio galaxies with redshifts \mbox{$z>4.5$}:
\mbox{$z=5.199$ \cite{debreuk_z:Verkhodanov2_n_en}} and $z=4.514$ \cite{kopylov_z:Verkhodanov2_n_en}
have been discovered using this very criterion (except one
discovered at $z=4.88$ with use of $L_\alpha$-line by Jarvis et
al. \cite{darvis:Verkhodanov2_n_en}). We cross-identified the selected radio
galaxies with the CATS radio catalogs, and computed the spectral
index at 325, 1400 and 4850\,MHz.

In Paper~I we as well constructed for our sample the ``spectral
index--redshift'' ($\alpha(z)$) relation, and other diagrams and
distributions.

Herein we report similar statistical data for the results of
optical observations initially catalogued in the NED and SDSS
databases. Note that photometric estimates of the parameters of
giant elliptical galaxies can be used for a number of cosmological
tests, such as the Hubble diagram ``magnitude--$z$''
 (K--$z$)
   \cite{willott_k_z:Verkhodanov2_n_en},
or the \mbox{``age--$z$''} diagram~\cite{starob:Verkhodanov2_n_en}. There are a
number of tests that allow the parameters and evolutionary
characteristics of the Universe to be estimated based on the radio
data (see, e.g.,~\cite{vo_par1:Verkhodanov2_n_en,vo_par2:Verkhodanov2_n_en,vo_par3:Verkhodanov2_n_en}).

As we pointed out above, this is the second paper of the planned
series of three papers dedicated to the construction of a sample,
and to the statistical analysis of the catalog of radio galaxies,
including the objects with spectroscopically measured $z$,
emitting at radio frequencies, and having optical photometric
data. In the future, we plan to use the physical parameters of the
catalog objects to perform the cosmological tests mentioned in
Paper~I~\cite{rg_listI:Verkhodanov2_n_en}.

\section{THE CATALOG}

\subsection{Description of the Catalog}

In this paper we describe the second part of the list of radio
galaxies, which contains optical magnitudes. The data are listed
in the Table presented in the Appendix. The columns of the table
give the names of the objects and their magnitudes. The complete
catalog is available from the CATS database\footnote{\tt ftp://cats.sao.ru/pub/CATS/RGLIST},
and CDS ftp\footnote{\tt ftp://cdsarc.u-strasbg.fr/pub/cats/J/other/AstBu/64.276},
CDS database\footnote{\tt http://cdsarc.u-strasbg.fr/cgi-bin/VizieR?-source=J/other/AstBu/64.276}.
In the table we use standard
designations for the passbands in which the magnitudes are
measured: the SDSS filters\footnote{The SDSS magnitudes are given
in the AB system as defined by Petrosian~\cite{sdss_cat:Verkhodanov2_n_en}.} u, g,
r, i, and z; the passbands of the Johnson and  UKIDSS systems: R,
U, B, V, G, H, I, J, K, R, and Z;
the Gunn system filters r$_G$ and i$_G$;
the Hubble Space Telescope's filters F160W, F775W, F850LP, F702W,
F606W and F814W; the ultraviolet filters FUV and NUV with the
1150--1700\,\AA~ and \mbox{1575--3110\,\AA} passbands,
respectively; and the Palomar Atlas photographic magnitudes of the
blue and red plates O and E.

The coordinates of the selected radio galaxies are listed in the
Table presented in the Appendix to Paper~I. Figure~1 shows the
positions in the sky for these objects.

\subsection{Statistical Analysis of the Sample}

We already pointed out in Paper~I that our list of galaxies is
neither complete in terms of sky coverage, nor homogeneous in
terms of sensitivity and frequency intervals, as the sample
contains objects drawn from different catalogs and unconnected sky
areas. Figure~1 demonstrates the positions of the selected
galaxies in the sky. The most complete and homogeneous subsample
of the list contains objects from the SDSS  survey, highlighted in
Fig.~1 with a different marker. The SDSS radio galaxies (the
NVSS~\cite{nvss:Verkhodanov2_n_en} and FIRST~\cite{first:Verkhodanov2_n_en} radio data) include many
objects with small redshifts ($z<0.5$) and small fluxes
\mbox{($S<15$\,mJy)}, making this subsample distinguish itself
quality-wise among other subsamples.

We performed a preliminary statistical analysis of our list and
present the results in Figs.~2 and 3. ~Figure~2 shows the
histograms of the magnitude distribution in the H, I, J, K, R, V,
g, i, r, u, and z passbands.

It is evident from this figure that the magnitudes of the objects
have different distributions in different filters. However, the
fact that for many radio galaxies there exist measurements in more
than three filters will make it possible in the future to estimate
the age of the stellar populations in these galaxies. In our
analysis of the sample we subdivided the sources into two groups:
(1) those with the semimajor axes  greater than $29''$, and (2)
those with the semimajor axes smaller than $29''$ (data adopted
from the NVSS catalog). Here $29''$ is the median semimajor axis
(according to NVSS) of the sources of our sample. In Figs.~3 and
4, the galaxies from groups (1) and (2) are marked with crosses
and circles, respectively. We made such division in order to
search for eventual physical differences between radio galaxies
depending on their radio sizes, what could be of help in the
studies of giant~\mbox{radio galaxies~\cite{grg_piter08:Verkhodanov2_n_en}.}
However, it is evident from the figures that the diagrams for the
subsamples selected based on this parameter differ
insignificantly.

\begin{table}[!tbp]
\caption{ {\bf Table.} Parameters of the
regression relation for the g, i, r, and z filters. Two
populations denoted by subscripts 1 and 2, are identified for the
g and z  passbands, respectively.}
\begin{tabular}{c|c|c}
\hline
 \quad Filter \quad \quad   &  \quad \quad  \quad $p$ \quad  \quad \quad  \quad & \quad \quad \quad $q$ \quad \quad  \quad   \\
\hline
 g$_1$     & -2.15 & 0.12    \\
 g$_2$     &  0.08 & 0.02    \\
 i         & -2.34 & 0.15    \\
 r         & -2.20 & 0.14    \\
 z$_1$     & -2.01 & 0.14    \\
 z$_2$     &  0.43 & 0.003   \\
\hline

\end{tabular}
\vspace{1cm}
\end{table}

The  g, i, r and z-band magnitudes show linear dependencies
$mag(z)$, which we derived based on the mean values averaged
inside the $\Delta z=0.5$ wide bins. The regression relation can
be described by the law $z = p + qx$, where $z$, $p$ and $q$ are
the redshift, intercept, and slope of the linear relation,
respectively. The Table above lists the parameters of the
regression relation for the g, i, r, and z filters. We see no
variation of redshift with magnitude changes in the u-band.
Moreover, two populations of objects characterized by different
regression relations can be identified in the g and z filters (see
Fig.\,5).

Below we list the radio galaxies that do not fit within the general relation in the Hubble diagrams:\\

\noindent 2MASX J15542080$+$2712295 ($z$=1.439),\linebreak TXS
0828$+$193 ($z$=2.572),\linebreak SDSS  J162626.28$+$371441.6
($z$=2.656),\linebreak SDSS  J110245.80$+$602830.6
($z$=2.111),\linebreak SDSS  \mbox{J143631.27$+$431124.7}
($z$=4.261),\linebreak SDSS  J082907.43$+$064546.0
($z$=2.224),\linebreak SDSS  J074324.07$+$233626.0
($z$=2.480),\linebreak SWIRE  J104528.29$+$591326.7
($z$=2.310).\linebreak

\section{CONCLUSIONS}

We report the second part of the radio galaxies catalog with
redshifts $z>0.3$, which contains complete information on the
magnitudes of the objects. The catalog is based on lists of
objects from the NED and CATS databases. Our sample includes the
SDSS objects, and the objects from the ``Big Trio'' program. We
perform a preliminary statistical analysis of the distributions of
individual magnitudes. The catalog will serve as a basis for
further studies of the ages of distant radio galaxies. We
constructed the Hubble diagrams for 11 filters, and determined the
parameters of linear regression for the g, i, r, and z filters of
the SDSS survey. In addition, two populations of objects
characterized by different regression relations can be identified
in the g- and z-band filters.

%
\noindent
{\small
{\bf Acknowledgments}.

This research has made use of the NASA/IPAC Extragalactic Database (NED) which is operated by the Jet Propulsion
Laboratory, California Institute of Technology, under contract with the National Aeronautics and Space
Administration. We also made use of the CATS database of radio astronomical catalogs
\cite{cats:Verkhodanov2_n_en,cats2:Verkhodanov2_n_en}
and  FADPS\footnote{\tt http://sed.sao.ru/$\sim$vo/fadps\_e.html}  radio-astronomical data reduction  system~\cite{fadps:Verkhodanov2_n_en,fadps2:Verkhodanov2_n_en}.
This work was supported by the Program of the Support of Leading Scientific Schools in Russia
(the School of S.\,M.\,Khaikin). O.V.V. acknowledges partial support from the Russian Foundation
for Basic Research (project \mbox{No\,07-02-01417-a})
}

\end{document}